\documentclass[twocolumn,showpacs,prc]{revtex4-1}

\usepackage{graphicx}
\usepackage{dcolumn}
\usepackage{bm}

\begin{document}

\title{Direct Evidence of Washing out of Nuclear Shell Effects \\}

\author{A. Chaudhuri} 
\author {T.K. Ghosh} \email{tilak@vecc.gov.in}
\author{K. Banerjee}
\author{S. Bhattacharya}
\author{Jhilam Sadhukhan}
\author{C. Bhattacharya}
\author {S. Kundu}
\author {J. K. Meena}
\author {G. Mukherjee}
\author {R. Pandey}
\author {T.K. Rana}
\author {P. Roy}
\author {T. Roy}
\author {V. Srivastava} 
\affiliation{Variable Energy Cyclotron Centre,  1/AF  Bidhan  Nagar,
  Kolkata  700 064, India}
\author{P. Bhattacharya}  \affiliation{Saha Institute of Nuclear Physics,  1/AF  Bidhan  Nagar,
  Kolkata  700 064, India.}

\date{\today}

\begin{abstract}
Constraining excitation energy at which nuclear shell effect washes out has important implications on the production of super heavy elements and many other fields of nuclear physics research. We report the fission fragment mass distribution in alpha induced reaction on an actinide target for wide excitation range in close energy interval and show direct evidence that nuclear shell effect washes out at excitation energy $\sim 40$ MeV. Calculation shows that second peak of the fission barrier also vanishes around similar excitation energy.

\end{abstract}

\pacs{25.70.Jj, 25.85.Ge}

\maketitle

One of the major areas that have generated unprecedented interest among contemporary nuclear physicists and chemists is the synthesis of super heavy elements (SHE). It is known from the Liquid Drop Model (LDM) of the nucleus \cite{Bohr} that if the two fundamental nuclear parameters, the attractive nuclear surface potential and the repulsive coulomb forces are taken into account, then our nuclear chart may end at around element number 104. This is simply because, nuclei with $Z\geq 104$ immediately fission as there is no barrier to prevent their decay. However, elements have been synthesized beyond that atomic number \cite{Oganessian}.

The observed stability of these heavy elements is believed to originate from the microscopic shell effects in nuclei. While LDM predicts the bulk properties of nuclei and explains their collective behavior, nuclear Shell Model \cite{Mayer} explains these shell gaps and the single-particle nature of nuclear states. Both the bulk properties and the shell properties of nuclei can be incorporated by adding a shell-correction term to the liquid-drop model energy.  Strutinsky \cite{Strutinsky, funny_hill} considered the shell effect as a deviation from uniform liquid drop model prediction and used the shell averaged single particle energy as a correction term to the liquid drop model energy. The liquid drop barrier height diminishes smoothly with the increase in atomic number as the nuclear fissility increases. However, as the shell correction term retains the fluctuations in the shell model energy, it is found that the incorporation of this shell correction alters the fission barrier and in fact, causes to develop large barrier to decay that can increase alpha or fission half-lives by several orders of magnitude for the heavy elements. Thus shell effects play a central role in determining the stability of the super heavy elements. Many important nuclear phenomena such as the fission isomers \cite{isomer_Strutinsky}, super deformed nuclei \cite{Robe} and new magic numbers in the exotic nuclei \cite{RituPRL} are the consequences of the shell effect.

It is generally believed that shell effects are washed out at higher excitation energy \cite{Vandenbosch}. For the production of the super heavy elements by heavy ion bombardment on actinides targets, the compound nuclei are always formed with an excitation energy exceeding a few tens of MeV. Judicious choice of the excitation energy is critical as the production cross section of the SHE may be increased by a few orders of magnitude if the beam energy is increased by few MeV. Therefore, constraining the excitation energy at which shell effects get washed out is really important in the context of the production of SHE. 

Fission fragment mass distribution (FFMD) of actinide nuclei has been studied in some detail by several authors \cite{back80, back81, colby}. In a radio-chemical study of fission fragments of alpha induced fission of $^{238}$U  Colby {\sl et al} \cite{colby} showed that the mass distribution in fission of $^{242}$Pu are asymmetric up to a lab energy of about 40 MeV, pointing to presence of the shell effect. Back {\sl et al} \cite{back81} showed that for $^{242}$Pu,  even at an excitation energy of 45-50 MeV, shell effect persists and the FFMD are asymmetric. But in 310 MeV $^{16}$O inelastic scattering on $^{238}$U, Back {\sl et al} \cite{back80} observed symmetric mass distributions at high excitations signifying washing out of the shell effects, and asymmetric mass distributions at low excitations. However, for a particular actinide element, the exact energies at which the shell effects disappear could not be found out in the above experiments. In this paper, we report the FFMD in alpha induced fusion-fission reaction on $^{232}$Th target at a wide excitation energy range  of 21-64 MeV and for the first time show direct evidence that the  shell effect is washed at excitation energy of about 40 MeV in $^{236}$U. 

The experiment was performed with $^4$He beam from the K-130 cyclotron at the Variable Energy Cyclotron Centre, Kolkata, India. The target was a self-supporting $^{232}$Th of thickness 1.1 $mg/cm^2$. For the detection of fission fragments, two large-area (20 cm $\times$ 6 cm) position-sensitive multiwire proportional counters (MWPCs) \cite{myNIM} were placed at the folding angle, covering 67$^{\circ}$ and 83$^{\circ}$, respectively, on either side of the beam axis. For each fission event, the time difference of the fast anode pulses of the detectors with respect to the pulsed beam, the X and Y positions together with the energy loss of fission fragments were measured. The operating pressure of the detectors were maintained at 3.0 torr of isobutane gas.  At this low pressure, the detectors were almost transparent to elastic and quasi-elastic particles. The polar angle of emitted fission fragments could be determined with accuracy better than 0.2$^{\circ}$ while the accuracy in azimuthal angle was about 0.8$^{\circ}$. Beam flux monitoring as well as normalization was performed using the elastic events collected by a silicon surface barrier detector placed at forward angle and the total charge collected at the Faraday cup. The event collection was triggered by the detection of a fission fragment in any of the MWPC detectors along with the beam pulsing of the Cyclotron. 

\begin{figure}
\includegraphics*[scale=0.4, angle=0]{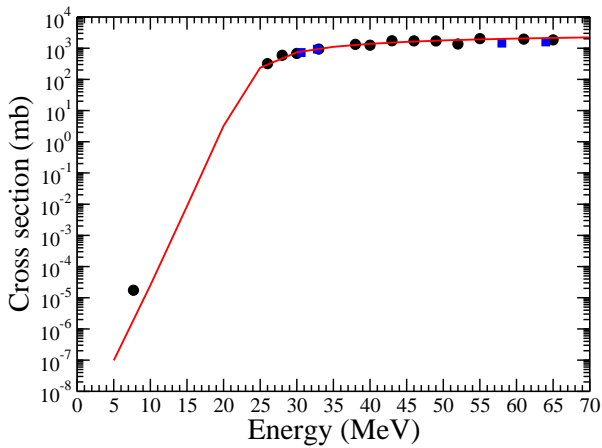}
\caption{\label{fig:fig1}~(Color online) The excitation function of fission for $^4$He+ $^{232}$Th reaction. The solid (black) circles are the present measurement. Measurement of  Ralarosy {\sl et al.} \cite{PRC73} is shown by solid (blue) squares.} 
\end{figure}

The fission fragments were well separated from quasi-elastic channels, in both coincident time and energy loss spectra. Fig. 1 shows the measured excitation function for fission.  The results of Ralarosy {\sl et al.} \cite{PRC73} , which are in agreement with the present measurement, are also shown in the figure. The systematic and statistical errors in the spectra are smaller than the size of the data points in the figure. The solid line in the figure corresponds to the coupled channel prediction (CCDEF) \cite{CCDEF}. In the present calculation, we have used axially symmetric shape of the target, characterized by nuclear quadruple and hexadecapole deformation parameters $\beta_2$ = 0.217 and $\beta_4$ = 0.09 \cite{PRC95nayana}. The agreement between the experimental and theoretical excitation functions is quite satisfactory at all energies around and above the barrier. However, the fission cross section measured in the same experiment at deep sub-barrier energy (7.7 MeV), where no measurement have been reported so far, shows enhancement compared to the theoretical prediction. It is worth pointing out that the measurement of fission cross section at deep sub-barrier energies is experimentally challenging and this phenomenon of enhancement of cross section is of particular interest for extreme sub-barrier fusion reactions of astrophysical interest \cite{JiangPRL02, AradhanaPRL09}. However, the mechanism of fission enhancement is not the central topic here and will be discussed in detail elsewhere \cite{Abhirup}.

Fig. 2 shows a typical distribution of the complementary fission fragments in ($\theta,\phi$) at excitation energy, E* = 23 MeV. The polar and azimuthal angle correlation for the fission fragments shows that the fission followed complete fusion  and formation of compound nucleus. The width of the polar and azimuthal angular correlations  includes, in addition to the spread due to fission reaction kinematics, the spread due to neutron emission from fragments. To avoid large  angular deviations due to neutron emissions washing out kinematic correlations of the complementary fission fragments, the experimental events within a  high intensity region in the middle of the $\theta$- $\phi$ correlation plot corresponding to an angular cone of radius 4$^o$, as shown in the figure (black circle), were analyzed for mass determination.  

\begin{figure}
\includegraphics*[scale=0.25, angle=0]{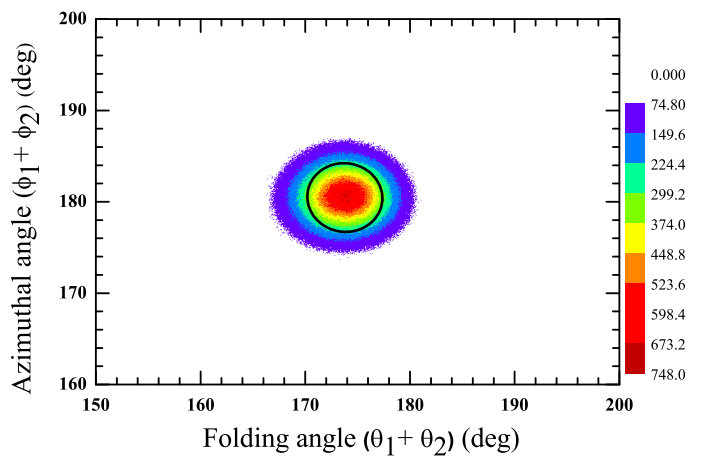}
\caption{\label{fig:fig2}~(Color online) Measured distributions of 
folding angles of the fissioning nuclei formed in the reaction $^{4}$He+$^{232}$Th  at an excitation energy of 23 MeV.} 
\end{figure}

The masses of the fission fragments were determined event by event from precise measurements of flight paths and flight time differences of the complimentary fission fragments \cite{myNIM,myPRC09}. The extracted FFMD at different excitation energies of the compound nucleus are shown in Fig. 3. Because of very low cross section, we could not measure the FFMD at alpha particle energy of 7.7 MeV (which was obtained in 3rd harmonic operation of the cyclotron). It is observed that, for excitation energies between 43.6 - 64.2 MeV (Fig. 3 i-l), mass distributions are symmetric in shape and are well described with a single Gaussian function peaking around approximately half of the mass of the compound nucleus. Since in these energies, quasi-fission is not expected, the symmetric mass distributions are more probable to originate from fission of a fully equilibrated compound nucleus through the saddle point along the macroscopic (LDM) barrier. Microscopic (shell) effect on the fission barrier is not significant at these energies. The widths of the FFMD would be determined by statistical process and would be a smooth function of temperature or excitation energy. On the other hand, it is noted that the shape of the FFMD changes from symmetric to asymmetric at excitation energies $\leq$ 40.5 MeV (Fig. 3 a-h).

\begin{widetext} 

\begin{figure}
\includegraphics*[scale=0.60, angle=0]{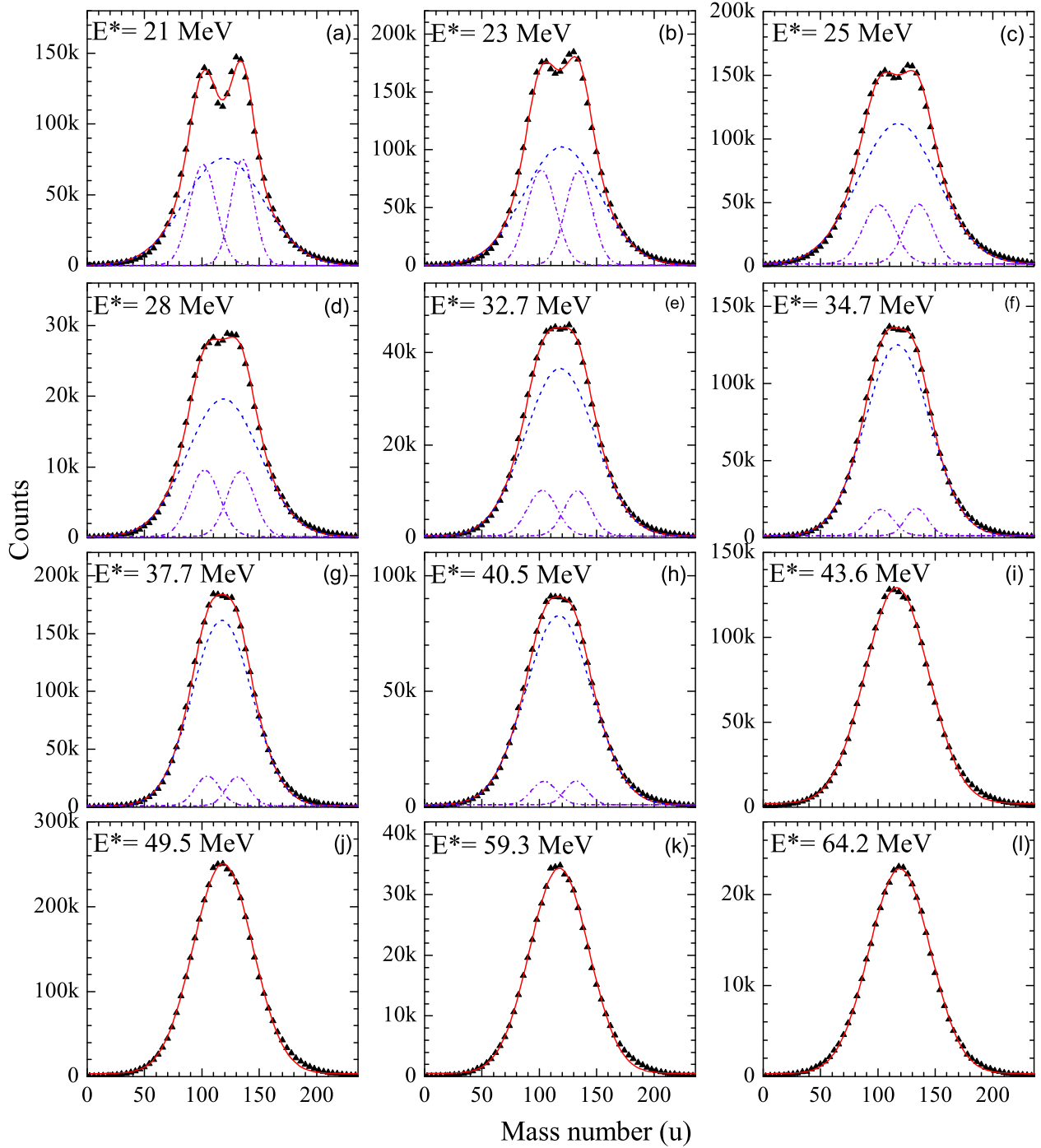}
\caption{\label{fig:fig3}~(Color online) FFMD at different excitation energies. Fitting by three Gaussians for E* = 21 MeV to 40.5 MeV are shown by dash blue (symmetric component) and violet dash-dotted (asymmetric components) lines. The overall fitting is shown by full (red) lines. The mass distributions at higher excitation energies ($\geq$ 43.6 MeV) are best fitted by a single Gaussian.} 
\end{figure}

\end{widetext}

The characteristic features of FFMD at lower excitation have been further elucidated in Fig. 4 using the data at 23 MeV excitation energy. The data cannot be fitted by a single symmetric Gaussian distribution, centered at symmetry. In the top half of the figure marked (a), we have tried to fit the data by two Gaussian functions of equal area which would be the scenario assuming asymmetric fission as observed in the case of spontaneous or thermal neutron induced fission \cite{Vandenbosch}. However, based on both the relative $\chi^{2}$ values and the visual inspection of the fits, it is found that the distribution could be best fitted by three Gaussians, with one peak corresponding to the symmetric division (A $\sim$ 118) and the other two at  A $\sim$ 132 and A $\sim$ 100 . This observation points to the co-existence of both asymmetric and symmetric fission at this excitation energy. However, the mass resolution of the set up and the procedure of fitting of three Gaussian are not adequate enough to pin point the reasons for the apparent asymmetry of the mass distribution with respect to half of the compound nuclear mass.

\begin{figure}
\includegraphics*[scale=0.4, angle=0]{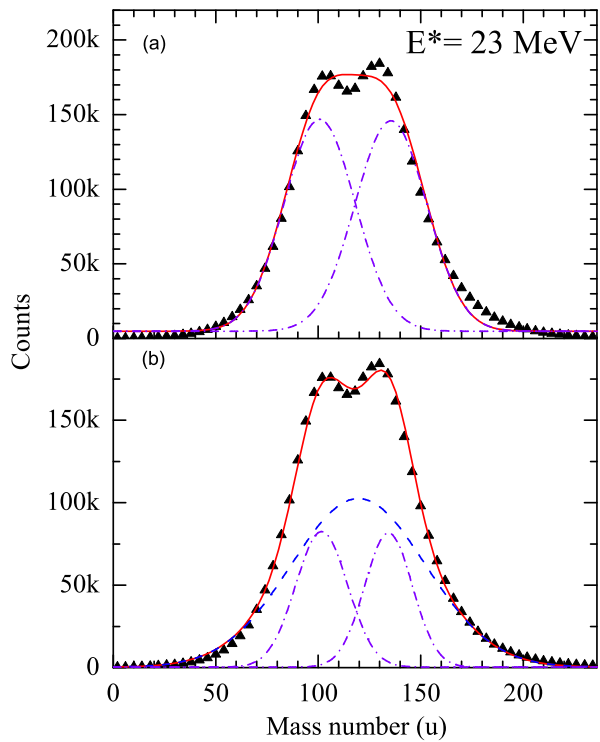}
\caption{\label{fig:fig4}~(Color online) FFMD at excitation energy 23 MeV fitted by two Gaussian (upper panel) and three Gaussian (lower panel) distributions. The asymmetric components are shown by (violet) dash-dot line and symmetric component is shown by (blue) dash line.The overall fitting is shown by solid (red) lines.} 
\end{figure}

It is interesting to note that all FFMDs with E* $\leq$ 40.5 MeV (Fig. 3 a-h) are best fitted with three Gaussian functions as in the case for E* = 23 MeV. However, there was a steady decrease in the total area under the Gaussians for asymmetric division, as the excitation energy increased; at 40.5 MeV, the two asymmetric peaks were barely discernible and then completely vanished at 43.6 MeV, where the experimental data could be fitted with a single Gaussian. Thus, symmetric mass fission is only mode present above 43.6 MeV. This can be viewed more quantitatively by looking at the ratio of the areas of the symmetric to the total yields ($G_{sym}$ / ($G_{sym} + G_{asy1} + G_{asy2}$), where $G_{sym}$, $G_{asy1}$ and $G_{asy2}$ are the areas under symmetric and two asymmetric components), plotted as a function of excitation energy as shown in Fig. 5. It is seen that the probability of fission from the symmetric mode increases as the excitation energy of the fissioning system is increased. At excitation energy $\sim$ 40 MeV, the value saturates to unity, clearly indicating the washing out of the asymmetric component of the mass distribution.

\begin{figure}
\includegraphics*[scale=0.3, angle=0]{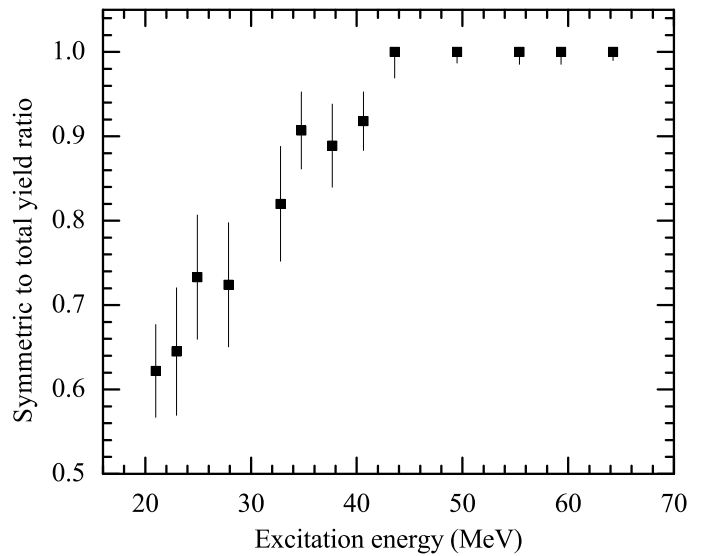}
\caption{\label{fig:fig5}~The variation of the ratio (relative unit) of the symmetric fission yield to the total fission yield at different excitation energies.} 
\end{figure}

Further insight of the fission process may be had by studying the widths of the FFMDs. It is well known that the width of the symmetric mass distribution is proportional to the temperature \cite{BackPRC96,myPLB} for the decay of hot statistically equilibrated compound nucleus. In Fig. 6, red dotted line shows the expected variation of the width of symmetric mass distribution with excitation energy. The black triangles represent the widths of the symmetric mass distributions as shown in Fig. 3. It is observed that only for energies above 43.6 MeV the data points follow the expected trend. Attempts were made to fit the mass distributions by constraining the width of the symmetric distribution around the expected trend (dotted line), the red solid square represents the fitted width.  It can be seen that such fitting is associated with very large uncertainty (shown by red vertical lines) and thus unphysical. It is observed that at excitation energy lower than 43.6 MeV, the best fitted width of the symmetric distribution (black triangles) increases with decrease in energy. This effect of increase in width of FFMD with decrease in excitation energy, may also be a signature of onset of shell effect. Such an effect has not been seen before and needs more detailed investigation.

\begin{figure}
\includegraphics*[scale=0.3, angle=0]{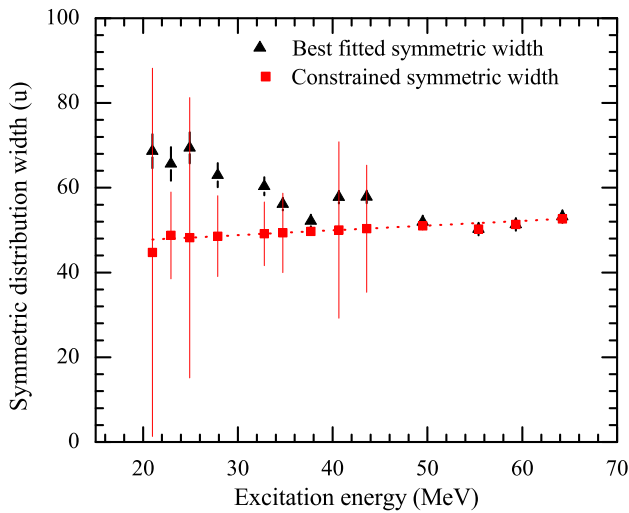}
\caption{\label{fig:fig6}~(Color online) Variation of the width of the fitted symmetric mass distribution with excitation energy. } 
\end{figure}

Thus it is clearly observed  that there is co-existence of two fission modes in $^4$He + $^{232}$Th reactions at low excitation energies, one leading to symmetric mass distribution and the other leading to a mixture of symmetric and asymmetric mass distributions. While the symmetric component can be explained by the liquid drop model, the asymmetric component is likely to be arising due to microscopic shell effects.Through shape oscillations, the compound nucleus has to passover a fission barrier which is described as a combination of macroscopic (LDM) and a microscopic (shell effect) barrier. The minimum energy path to scission would be a statistical mixture of probabilities in which the mass distribution could be decided at LDM (symmetric) or the LDM plus shell corrected (asymmetric) fission barrier. The present experiment shows that at lower excitation energies these two fission modes co-exist, but the asymmetric component gradually vanishes at around 40 MeV, as evident from the mass distributions. We consider the vanishing of the asymmetric mode of fission around 40 MeV as a direct signature of the washing out of shell effect in $^{236}$U.  

In spontaneous fission and in many reactions involving fission of actinides and pre-actinide nuclei, a distinct asymmetry in FFMD was seen \cite{Vandenbosch} at lower excitation energies. Within the framework of the shell correction method proposed by Strutinsky \cite{isomer_Strutinsky}, nuclear potential is obtained from the superposition of a macroscopic smooth liquid drop part and a shell correction term, obtained from microscopic single particle model. As a result, for heavy nuclei like $^{236}$U, the potential shows the double-humped character as a function of deformation. Potential energy surface calculation shows \cite{MollerNature01} that the saddle point corresponding to second barrier has a mass-asymmetric shape for heavy nuclei. Thus FFMD should be asymmetric if the fragment passes over the shell corrected potential. We have observed that the position of the heavier peak around 132 - 134 which is in the vicinity of doubly shell magic $^{132}$Sn nuclei. It is generally observed that in the fission of actinides nuclei, the constancy of the heavy mass fragments occurs around mass number 140 due to deformed shell. However, in the absence of measurement of total kinetic energy of fission fragments, we could not be certain about the above effect.

\begin{figure}
\includegraphics*[scale=0.5, angle=0]{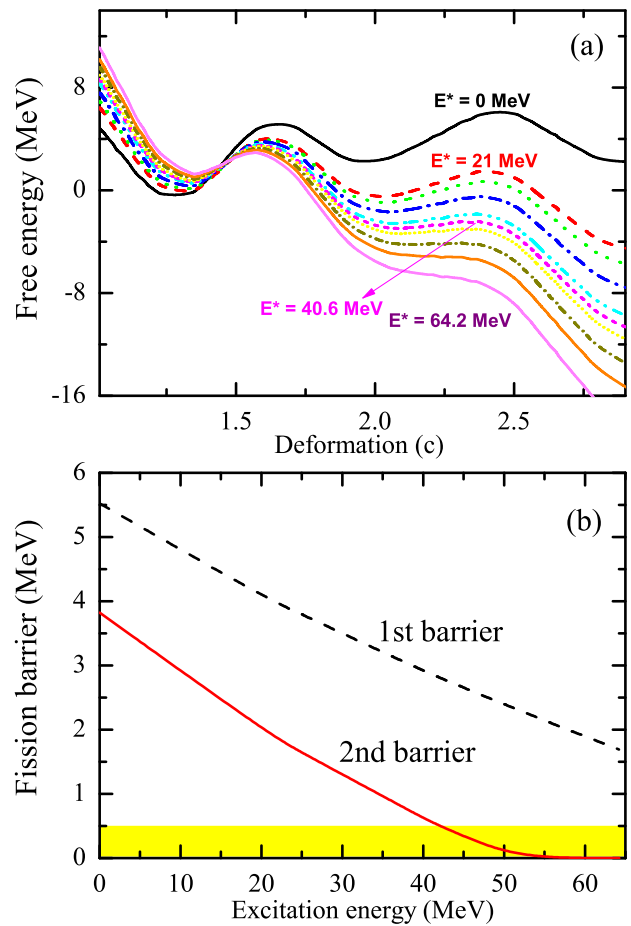}
\caption{\label{fig:fig7}~(Color online) (a): Variation of the free energy as a function of deformation ($c$, see text) for $^{236}U$ for ground state and excitation energies 21, 24.9, 30.8, 37.7, 40.6, 43.6, 49.5, 55.3 and 64.2 MeV. (b): Variation of fission barrier as a function of excitation energy. The shaded (yellow) region represents the uncertainty in fission barrier calculation.} 
\end{figure}

It is evident that the nature of variation of fission mode (symmetric to total yield ratio) will critically depend on the nature of variation of the corresponding barriers with excitation energy. The variation of this barrier with excitation energy may be understood from the nuclear free energy $F$ that determines the collective dynamics of a hot compound system \cite{nadtochyPRC02,lestonePRC09}. The expression for free energy is given by $F=$V$-(a-a_{gs})T^2$, where $V$ is the potential energy, $a$ is shape-dependent level density parameter with the value $a_{gs}$ at the ground state deformation. The nuclear temperature $T$ is calculated at the ground state deformation. Following the Fermi gas model, $T$ can be obtained from the intrinsic excitation energy $E^*$ by using the relation: $E^*=a_{gs}T^2$ \cite{bohr_mottleson}. In the present calculation, nuclear shapes are defined within the ellipsoidal shape parametrization, where $c$, the ratio of the symmetry axis to any other principal axis of the ellipsoid, quantifies the amount of deformation. We have used the shell corrected $V$  obtained from a macroscopic-microscopic model \cite{PLB}. The value of $a$ is calculated following the work of Ignatyuk \textsl{et al.,} \cite{ignatyuk1975,jhilam_prc2008}.

The variation of $F$ as a function of deformation $c$ of the system is plotted in Fig. 7 (a) for different values of $E^*$. For the system $^{236}$U, it is evident form Fig. 7 that, there exits two different fission barriers separated by a second minima. It is clear from Fig. 7 (a) that the heights of the two fission barriers decrease with $E^*$. The variation of the fission-barrier height  as a function of $E^*$ is shown in the lower panel (b) of the figure. It is seen that the second barrier merely vanishes (less than 500 KeV) at around $\sim 40$ MeV. It is interesting to note that in our measurement, the asymmetric fission fragment yield also vanishes at same excitation energy. So, it can be inferred that the observed vanishing of asymmetric mass yield is correlated with the vanishing of the second peak of the double hump fission barrier and vis-a-vis the vanishing of shell effect. 

As mentioned earlier, nuclear shell effect also affects the nuclear level density (NLD). From fission fragment angular distribution, it was shown \cite{PRL70Ramamurthy} that the shell effect on nuclear level density parameter would be damped with excitation energy so that the level density parameter value reaches its liquid drop value at around  the similar excitation energy ($\sim$ 40 MeV) where we find the asymmetric component of the mass distribution vanishes. From the measured proton evaporation spectra \cite{EPJ02} in nuclei around $^{208}$Pb at $E^*$ $\sim$ 50 MeV, the extracted NLD also showed the expected liquid drop behavior. The present data, therefore, are consistent with the above findings.

It may be worthwhile to mention here that the measurement of masses of the fission fragments, in the present technique is less susceptible to be modified by secondary de-excitation of excited fission fragments as the mean fragment velocity (measured here), unlike the kinetic energy, does not change due to the particle evaporation. We have also rejected the events where the flight paths are greatly modified by neutron evaporation. 
 Moreover, as the angular momentum involved in $\alpha$ induced reaction is much less, the effect of angular momentum dependence of fission barrier does not significantly affect the barrier to modify the results. As fusion fission reaction was chosen here, excitation energy estimation is also less ambiguous. Also, unlike in heavy ion induced fusion,  $\alpha$ induced fusion is completely free from other competing processes (e.g; quasi-fission) which could otherwise contaminate the mass distribution. Most importantly, for the present experiment, the extracted results are completely model independent; on the contrary, the results derived from either angular anisotropy \cite{PRL70Ramamurthy} or proton/gamma evaporation studies \cite{EPJ02} require specific model calculation to extract information though the weakness of shell effect with increase in excitation energy was known qualititively.

In conclusion, though the weakening of shell effect with increase in excitation energy was known qualitatively from previous studies, for the first time, we show a direct evidence that nuclear shell effect gets washed out at  $E^*$$\sim$ 40 MeV. The asymmetry in mass distribution observed in our experiment, at lower excitation energies is due to the manifestation of shell effects. From the FFMD, it is clear that the symmetric distribution component increases with the increase in excitation energy, indicating that shell effects are more prominent at lower excitation energies. The change in shape of the mass distribution, from asymmetric to symmetric, at  $E^*$$\sim$ 40 MeV is a direct evidence of the washing out of shell effects. A systematic study along this line for other actinide elements should be carried out to understand the role of nuclear shell effect in a better way. 

We are thankful to P.S. Chakraborty and staff members of the K130 Cyclotron at Variable Energy Cyclotron Centre for providing good quality alpha beam required for the experiment. We acknowledge the help of Kouichi Hagino for the coupled channel calculation. Thanks are also due to S.S. Kapoor, V.S. Ramamurthy and Santanu Pal for fruitful discussions.

\end{document}